\documentstyle[11pt,twoside,fleqn,espcrc2]{article}

\newcommand{\AmS}{{\protect\the\textfont2
  A\kern-.1667em\lower.5ex\hbox{M}\kern-.125emS}}

\title{PRESENT AND FUTURE OSCILLATION EXPERIMENTS AT REACTORS}

\author{L. Mikaelyan\address{Kurchatov Institute, Moscow, Russia}%
}
\begin{document}

\begin{abstract}
This is a report on recent progress and developments since the NANP'99 Conference in
current and future long baseline ($\sim$1 km) and very long baseline ($\sim$100 - 800 km) 
oscillation experiments at reactors. These experiments, under certain assumptions, 
can fully reconstruct the internal mass structure of the electron neutrino and 
provide laboratory test of solar and atmospheric neutrino problems.
\end{abstract}

\maketitle

\section{INTRODUCTION}

In this report we discuss the following experiments and projects:
\begin{itemize}
\item The first long baseline experiment at CHOOZ (France, Italy, Russia, US)[1] 
completed soon after NANP'99: final results 
\item The long baseline Palo Verde experi-ment in Arizona [2]: \ current results. 
The experiment is scheduled to be finished in 2000 y.
\item The long baseline two detector project Kr2Det for the Krasnoyarsk underground 
(600 mwe) site aimed to search for very small mixing angles [3]. The project 
is in a R$\&$D stage.
\item The very long baseline experiment KamLAND at Kamioka (Japan, the USA) [4]. \ The data 
taking can start in 2001 y.
\end{itemize}

To safe the pages we do not consider the very long baseline experiment BOREXINO at 
Gran Sasso, which has been reported in detail at this Conference by T. Hagner. 
We also refer to the reports given here by V. Sinev, who considers possibilities of 
testing the LSND oscillations in a reactor experiment, and V. Kopeikin who presents 
new information on the reactor antineutrino energy spectra important for data 
analysis.

We consider the oscillation experiments at reactors as an effective tool to 
investigate the internal mass-structure of the electron neutrino. In this we use 
analysis developed in [5].

All oscillation experiments considered here are based on the reaction of the inverse 
beta decay 
\begin{equation}
\qquad \qquad \bar{\nu} + p \rightarrow n + e^{+}
\end{equation}

with a threshold of 1.80 MeV, and use ($e^{+},n$) delay coincidence technique. 
The energy of the ejected positron is
$$
 T = E - 1.80 \quad MeV 	\qquad  \eqno (1')
$$
($E$ is the energy of the incoming anti-neutrino). In most cases the annihilation quanta 
are absorbed in the fiducial volume and the visible energy of positron is $\sim$1 \ MeV 
higher than that of Eq. (1). The first two experiments make use of Gd loaded liquid 
scintillator as a $\bar{{\nu}_{e}}$ target, for Kr2Det and very long baseline experiments no Gd is 
planned.

\section{Motivations}

2.1. First we remind of parameters that describe the oscillation process in the two-neutrino 
model. The survival probability $P(\bar{{\nu}_{e}}\rightarrow \bar{{\nu}_{e}})$ that the electron antineutrino will retain its 
initial flavor at a distance of $L$ meters from the source is given by the expression:
\begin{eqnarray}
 \qquad P(\bar{{\nu}_{e}}\rightarrow \bar{{\nu}_{e}})=1 - \sin^{2}2{\theta}\times \nonumber\\
\sin^{2}(\frac{1.27{\Delta}m^{2}L}{E})
\end{eqnarray}
where $\sin^{2}2{\theta}$ is the mixing parameter, ${\Delta}m^{2} (eV^{2}) \equiv m^{2}_{2} - m^{2}_{1}$ 
is the mass parameter and $m_{1,2}$ are masses of the interfering states.

A specific deformation of the measured positron (antineutrino) energy spectrum and a 
deficit of the total antineutrino detection rate relative to the no-oscillation case are 
the oscillation signatures, which are searched for in the experiment. Calculation shows 
that for soft reactor neutrinos the distortion of the energy spectrum is most pronounced 
and the deficit of the neutrino detection rate is maximal for 
\begin{eqnarray}
\qquad {\Delta}m^{2} L \approx 5 \ (eV^{2} m)  \nonumber \\
 \qquad (sensitivity \ condition)
\end{eqnarray}

2.2. No theory can predict today neutrino mass and mixing parameters. Positive information on 
this subject comes from the studies of the atmospheric and solar neutrinos.

The Super-Kamiokande observations of atmospheric neutrinos provide a strong evidence for 
neutrino oscillations. Recent data reported at NEUTRINO'2000, if interpreted as 
${\nu}_{\mu} \leftrightarrow {\nu}_{\tau}$ transitions, are best fit by the oscillation 
parameters [6]:
\begin{eqnarray}
 {\Delta}m^{2}_{atm}  \approx 3\times 10^{-3} \ eV^{2} (most  \nonumber \\
probable \ value), \ \sin^{2}2{\theta}_{atm} > 0.88,
\end{eqnarray}
It should be emphasized however that analysis of the atmospheric neutrinos leaves quite a 
large room for the ${\nu}_{\mu} \leftrightarrow {\nu}_{e}$ channel as a subdominant one.
	
The energy dependent deficit of the solar neutrinos relative to the Standard Solar Model 
prediction is another strong argument in favor of the oscillation hypothesis. By assigning 
particular values to the parameters ${\Delta}m^{2}$ and $\sin^{2}2{\theta}$ and by inclusion of the MSW mechanism all 
observations can be accounted for [7]. The most recent data from Super-Kamiokande as 
analyzed in [8] give however a strong preference to only one of the solutions labeled as 
Large Mixing Angle (LMA) MSW solution:
\begin{eqnarray}
\qquad {\Delta}m^{2}_{sol}  \approx 3\times 10^{-5} \ eV^{2}, \nonumber \\
\sin^{2}2{\theta}_{sol} \approx 0.7,
\end{eqnarray}

Other possibilities, the SMA MSW and vacuum oscillation solutions, are now strongly 
disfavored.

Eqs (3)-(5) show that the long baseline (LBL) and very long baseline (VLBL) reactor 
experiments explore the electron neutrino mixing respectively in the atmospheric and solar 
neutrino mass parameter regions.

2.3. We go now to the three active neutrino oscillations. In this case the mixing para-meters 
$\sin^{2}2{\theta}$ are expressed through the elements of the neutrino mixing matrix 
$U_{ei}$, which represent the contributions of the mass states to the electron neutrino 
flavor state ${\nu}_{e}$:
\begin{eqnarray}
\qquad {\nu}_{e} = U_{e1}{\nu}_{1} + U_{e2}{\nu}_{2} + U_{e3}{\nu}_{3};\nonumber\\
U^{2}_{e1} +U^{2}_{e2} + U^{2}_{e3} = 1,\nonumber\\
\sin^{2}2{\theta}_{LBL} = 4 U^{2}_{e3}(1 - U^{2}_{e3}),\nonumber\\
\sin^{2}2{\theta}_{VLBL} = 4 U^{2}_{e1} U^{2}_{e2}
\end{eqnarray}
(${\nu}_{i}$  are the mass eigenstates).

We conclude that the LBL and VLBL experiments at reactors can provide full information on 
the electron neutrino mass structure, at least in the 3-neutrino mixing model. It is 
interesting to mention that sensitive measurements of $U_{e3}$ can help to choose between 
possible oscillation solutions of the solar neutrino problem independent of the VLBL 
experiments [9].

\section{The CHOOZ Experiment}

The CHOOZ detector was built in an underground gallery (300 mwe) at distances of about 
1000 m and 1110 m from two RWR reactors of total nominal power 8.5 GW (th). The detector 
shown in Fig. 1 has three concentric zones. The central zone with 5 tons of hydrogen-rich 
Gd-loaded liquid scintillator served as a target for antineutrinos. The target is immersed 
in a 70 cm thick intermediate zone filled with a Gd-free scintillator (17 tons) and 
surrounded by a veto outer region with 90 tons of ordinary scintillator. The inner two 
zones are viewed by 192 eight-inch PMT's.

The data was obtained at different power levels of the CHOOZ NPP newly built reactors 
as they were gradually brought into operation. This schedule was very useful for 
determining the reactor OFF background. A summary of data taking periods from April 1997 
till July 1998 is shown in Table 1.

\begin{table*}[hbt]
\setlength{\tabcolsep}{1.5pc}
\newlength{\digitwidth} \settowidth{\digitwidth}{\rm 0}
\catcode`?=\active \def?{\kern\digitwidth}
\caption{CHOOZ data acquisition periods}
\label{tab:aquis}
\begin{tabular*}{\textwidth}{@{}l@{\extracolsep{\fill}}cc}
\hline
 & Time (d) & W (GW) \\
\hline
Reactor 1 ON & 85.7 &  4.03  \\
Reactor 2 ON & 49.5 &  3.48  \\
Reactor 1 and 2 ON  & 64.3 & 5.72 \\
Reactor 1 and 2 OFF & 142.5  & $-$ \\
\hline
\end{tabular*}
\end{table*}
  
The selection of neutrino events is based on the following conditions: (i) energy cuts on 
the positron candidate (1.3 - 8 MeV) and on the neutron candidate (6 - 12 MeV), (ii) a time 
window on the delay between $e^+$ and neutron (2 - 100) ms, (iii) spatial cuts on the positron 
and neutron positions (distance from the PMT surface $>$ 0.3 m and distance between positron 
and neutron events $<$ 1.0 m). Under these conditions the antineutrino detection efficiency 
was found to be ${\epsilon} = (69.8 \pm 1.1)\%$.

Total about 2500 $\bar{{\nu}_{e}}$ were detected during the data acquisition periods. The measured 
neutrino detection rate is 2.58 per day per GW of reactor power and the typical ratio of 
the neutrino to background detection rates is 10:1. The ratio $R_{meas/calc}$ of the measured to 
expected for no-oscillation case neutrino detection rates is found to be:
\begin{eqnarray}
\quad R_{meas/calc}= 1.01 \pm 2.8\%(stat) \nonumber \\
\quad \pm 2.7\%(syst),
\end{eqnarray}
Ratio (7) was computed with \ the use \ of the reaction (1) cross section accurately 
measured by the KURCHATOV-IN2P3 group at a distance of 15 m from the Bugey-5 reactor [10]. 
Uncertainties which build up the total systematic error given in Eq. (7) are listed in the 
Table 2.

\begin{table*}[hbt]
\setlength{\tabcolsep}{1.5pc}
\catcode`?=\active \def?{\kern\digitwidth}
\caption{Components of the CHOOZ 68\% CL systematic uncertainties}
\label{tab:choozdata}
\begin{tabular*}{\textwidth}{@{}l@{\extracolsep{\fill}}c}
\hline
Parameter & Relative error \% \\
\hline
Reaction cross section & 1.9 \\
Number of protons      & 0.8 \\
Detection efficiency   & 1.4 \\
Reactor power          & 0.7 \\
Energy absorbed per fission & 0.6 \\
\hline
Combined               & 2.7 \\
\hline
\end{tabular*}
\end{table*}
The positron and background spectra measured during reactor ON and OFF periods are shown 
in Fig. 2a, the ratio of the measured to the expected spectrum can be seen in Fig. 2b. 
Clearly, neither the $\bar{{\nu}_e}$ detection rate, nor positron spectrum shows any signs 
of neutrino oscillation. 

The CHOOZ oscillation constraints are derived by comparing all the available experimental 
information to expected no-oscillation values. The result (Fig. 3 the curve ``CHOOZ'') 
directly depends on the correct determination of the absolute $\bar{{\nu}_e}$ flux and 
their energy spectra, the nuclear fuel burn up effects, $\bar{{\nu}_e}$ cross section, 
detector efficiency and spectral response$\ldots$

We note that CHOOZ does not observe the $\bar{{\nu}_e}$ oscillation in the mass region 
${\Delta}m^{2}_{atm}$ where muon neutrinos oscillate intensively: 
\begin{eqnarray}
\qquad \qquad \sin^{2}2{\theta}_{CHOOZ} \le 0.1, \nonumber \\
U^{2}_{e3} \le 2.5\times 10^{-2} \nonumber \\
\ (at \ {\Delta}m^{2} = 3\times 10^{-3})
\end{eqnarray}

The CHOOZ experiment has demonstrated a considerable improvement on the reactor $\bar{{\nu}_e}$
detection techniques: the level of the background at CHOOZ ($\sim$0.3 per day, per target 
ton) is almost a thousand times lower than has ever been achieved in previous experiments. 
In this connection we would mention two important points. The first is the underground 
position of the detector. The 300 mwe rock overburden reduces the flux of cosmic muons, 
the main source of the time-correlated background, by a factor of $\sim$300 to a value of 
0.4 m$^{-2}$ s$^{-1}$. The second point is associated with the zone-2 of the detector. 
The scintillator of this zone absorbs the radiation coming from high natural radioactivity 
of the PMT's glass and relevant events are rejected by the spatial cuts thus reducing 
the accidentals. These two features are specific for future LBL and VLBL projects with the 
difference that greater baselines require deeper detector positions and the protective 
region between the fiducial volume and PMTs is thicker and is filled with non-scintillating 
mineral oil.

\section{The Palo Verde Experiment}

This experiment uses detector of quite a different design and more sophisticated selection 
criteria. The difference is caused by a shallow position of the laboratory (32 mwe) and 
$\sim$50 times higher muon flux than at the CHOOZ site.

The $\bar{{\nu}_e}$ target is a $6\times 11$ matrix composed of 12.7 cm $\times$ 25 cm 
$\times$ 900 cm acrylic cells. The inner 7.4 m-long part of the sell is active and 0.8 m 
on each side serve as oil light guides 
and buffers, which shield the central part from external radioactivity (Fig. 4). The total 
volume of the liquid scintillator (Gd) amounts to 12 m$^3$. A 1-m thick layer of purified 
water passive shielding surrounds the central detector. The outmost layer of the detector is 
composed of veto counters. The veto rate is typically $\sim$2 kHz.

The experiment is situated in Arizona, the USA. Three identical PWR type reactors of total power 
of 11.6 GW (th) are located at distances of 890, 890 and 750 m from the detector. Each 
reactor is shut down for refueling every year. Two of the reactors are ON at any given time.

The positron trigger is a fast (30 ns) triple coincidence between neighboring cells 
requiring one cell above 600 keV (positron ionization) and two cells above about 40 keV 
to detect Compton recoils from annihilation quanta. Similar conditions are used for the 
neutron candidates. The time delay between the positron and neutron ``triples'' was chosen 
about 450ms long, much longer than neutron capture time in the Gd-loaded scintillator 
($\sim$30 ms), which was useful for determining the accidental coincidence background. For 
details of subsequent cuts applied in the offline data treating we refer to Ref [2]. 

Presently (July 2000) are available results based on data taking period from July 1998 to 
September 1999. In 1998 one of the reactors at 890 m was OFF for 31 days and in 1999 the 
reactor at 750 m was OFF for about 23 days. The three reactors ON minus two reactors ON 
give the following neutrino detection rates (per day):
\begin{eqnarray}
\qquad 6.0 \pm 1.4 (stat) \ in \ 1998 \ and \nonumber \\
9.0 \pm 1.6(stat) \ in \ 1999. 
\end{eqnarray}
The neutrino detection efficiencies are estimated as 7.6\% (1998) and 11\% (1999). 
The rates are found to be compatible with no-oscillation predictions.

The ON-OFF method treats the $\bar{{\nu}_e}$ flux from the two reactors still at full power as 
background, which considerably reduces the statistical accuracy of the results. An 
independent analysis named the ``swap'' method is based on (i) the symmetry of the most of 
the backgrounds relative to the exchange of the first and the second subevents and on (ii)
 the strong asymmetry of the positron and delayed neutron signals. The ``swap'' analysis uses
 full neutrino statistics. It makes it possible to cancel most of the background directly 
from the data. The remaining part is computed using Monte Carlo simulations. 
The ratio $R_{meas/calc}$ of the measured to expected for no-oscillation case neutrino detection 
rates found by means of the ``swap'' method is:
\begin{eqnarray}
\quad R_{meas/calc}= 1.04 \pm 3\%(stat) \nonumber \\
\quad \pm 8\%(syst),
\end{eqnarray}
Clearly the gain in statistic relative to the classic ON-OFF method (Eq. 9) is quite 
impressive. The systematic uncertainties are summarized in Table 3.
\begin{table*}[hbt]
\setlength{\tabcolsep}{1.5pc}
\catcode`?=\active \def?{\kern\digitwidth}
\caption{The Palo Verde systematic uncertainties [2]}
\label{tab:pverde}
\begin{tabular*}{\textwidth}{@{}l@{\extracolsep{\fill}}cc}
\hline
Error source & On minus OFF (\% )& Swap (\%) \\
\hline
 $e^+$ efficiency & 4 & 4 \\
n efficiency      & 3 & 3 \\
neutrino flux prediction  & 3 & 3 \\
neutrino selection cuts   & 8 & 4 \\
BKG estimate  & $-$ & 6 \\
\hline
Total      & 10   & 8 \\
\hline
\end{tabular*}
\end{table*}
The best of the Palo Verde oscillation exclusion plot is shown in Fig. 3.

\section{The Krasnoyarsk: Two-Detector Project Kr2Det}

The Kr2Det $\sim$1 km baseline project is aimed at more sensitive searches for neutrino 
oscillations in the mass parameter region already studied in the CHOOZ and Palo Verde 
experiments. The physical goals of the project are: (i) to obtain new information on the 
electron neutrino mass structure ($U_{e3}$), (ii) to provide a better normalization for future 
accelerator neutrino experiments, and (iii) to achieve better understanding of the 
atmospheric neutrinos. It is also worth mentioning that just in case the LMA solution is 
valid $U_{e3}$ can be quite close to the presently available CHOOZ upper limit 
$U^{2}_{e3} \le 2.5\times 10^{-2}$ 
while for the SMA MSW and vacuum oscillation solutions the predicted value of $U_{e3}$ is much 
smaller [9]. 
The project intents: 
\begin{itemize}
\item To increase, relative to CHOOZ, the sample of detected neutrinos by a factor of 20. 
The Kr2Det neutrino target mass is 50 tons, ten times larger than used in the CHOOZ
experiment,
\item To eliminate most of the systematic uncertainties by using two identically designed 
scintillation spectrometer (far and near) stationed at 1100 m and 250 m from the reactor,
\item To use special calibrations to control and correct for systematic uncertainties that 
will still remain.
\end{itemize}

The detectors are installed at a depth of 600 mwe with the flux of cosmic muons there 5 
times lower than at the CHOOZ laboratory, which helps to keep the backgrounds at 
sufficiently low level.

Each of the detectors (Fig. 5) has a three-concentric zone design: the 50-ton liquid 
scintillator (no Gd) target in the center, the buffer of non-scintillating oil and the outer 
veto zone. Expected neutrino detection and background rates are: $N_{\nu} = 50 \ d^{-1}$, 
$N_{BKG} = 5 \ d^{-1}$.

In the no-oscillation case the ratio of the two simultaneously measured positron spectra 
does not depend on the positron energy, small deviation from the constant value is analyzed 
for the oscillation parameters. The results of this purely relative analysis are independent 
of the exact knowledge of reactor power and the fuel burn up effects, of numbers of target 
protons and detection efficiencie$\ldots$ Calibration of the detectors is a key problem of the 
experiment. More details on calibration procedures are considered in Ref. [2]

Expected 90\% CL oscillation limits are presented in Fig. 3. It was assumed that 40 
thousand $\bar{{\nu}_e}$ are detected and that systematics is controlled down to a 0.5\% level.

\section{KamLAND}

The KamLAND detector will operate in the Kamiokande detector cave with a rock overburden  
of 2700 mwe. The neutrinos originate from 16 NPP (total 51 power reactors) at distances 
of 80 km - 820 km. 80\% of the total $\bar{{\nu}_{e}}$ flux comes from reactors between 140 km and 210 km 
away. 

\begin{table*}[hbt]
\setlength{\tabcolsep}{1.5pc}
\catcode`?=\active \def?{\kern\digitwidth}
\caption{Some parameters of the underground oscillation experiments at reactors}
\label{tab:kaml}
\begin{tabular*}{\textwidth}{@{}l@{\extracolsep{\fill}}ccccc}
\hline
Parameter & L, km & mwe & target mass, t & $N_{\nu}$, t$^{-1}$ y$^{-1}$ & $N_{BKG}$, t$^{-1}$ y$^{-1}$ \\
\hline
CHOOZ  & 1.1 & 300 & 5 & 900 & 90 \\
Kr2Det$^{*)}$ & 1.1 & 600 & 50 & 370 & 40$^{**)}$ \\
KamLand & $\sim$200 & 2700 & 1000 & 0.8 & 0.08$^{**)}$ \\
\hline
\multicolumn{5}{l}{$^*)$ Detector in the 1100 m position} & \\
\multicolumn{5}{l}{$^{**})$ Estimated values } & \\
\end{tabular*}
\end{table*}

The reactor neutrino flux on the target is extremely small, 1000 times smaller than 
in the CHOOZ experiment, while the muon flux at this depth is attenuated only by a 
factor of $\sim$400 with respect to the value at the CHOOZ laboratory. Thus the background 
problems are of primary importance.

The KamLAND detector again has three concentric zones (Fig. 6). The 13-m diameter 
target zone with 1000 tons of purified scintillator is in the center. A 1700-ton buffer 
is filled with mineral oil and this time is 2.5 m thick. The scintillator is viewed 
by an array of 2000  photomultipliers supported on a 19-m diameter steel sphere. 
The zone-3 outside the steel sphere is filled with purified water and serves as a 
Cerencov veto detector and additional passive shielding. 

The calculated average neutrino detection rate is about 750 per year and is expected to 
vary by $\sim \pm$ 10\% due to the seasonal variation of the nuclear power production. 
Clearly the ON - OFF approach does not seem promising in this case. On the other 
hand the correlated background rates are estimated as low as about 20 events per 
year and the expected oscillation effect for LMA solar solution is quite large$\ldots$

The derived sensitivity to the oscillation parameters assuming three years of data 
taking and the ne to background ratio 10:1 is shown in Fig. 3. It can be seen that the 
LMA solution can be conclusively tested.

\section{Summary and Conclusions}

The $\sim$1 km baseline CHOOZ and, with a small delay, Palo Verde are the first of the 
terrestrial neutrino oscillation experiments that have successfully explored the 
atmospheric mass parameter region. The negative result of these experiments has an 
important positive meaning that the electron neutrino contains not much of the mass-3.

The success of the CHOOZ experiment is based on impressive (almost three orders of 
magnitude) improvements on the reactor neutrino detection techniques. 
A revolutionary progress in the field is well underway as can be seen from Tab. 4, 
which summarizes some parameters of the experiments discussed in this report.

More sensitive searches for the mass-3 admixture are feasible now (Kr2Det). An 
invasion into the LMA MSW solar region (KamLAND) requires another three orders of 
magnitude reduction of the backgrounds, which most probably can be achieved. 
We conclude that, with reactor experiments, in a few years the electron neutrino 
mass structure can be understood and a decisive proof of the solar neutrino 
problem found.

\section*{Acknowledgments}

I would like to thank the Conference organizers for an excellent meeting and generous 
hospitality. I am grateful to Dr. V. Sinev for assistance. This work is supported by 
RFBR grants Numbers 00-02-16035 and 00-15-96708.

\section*{Figures}
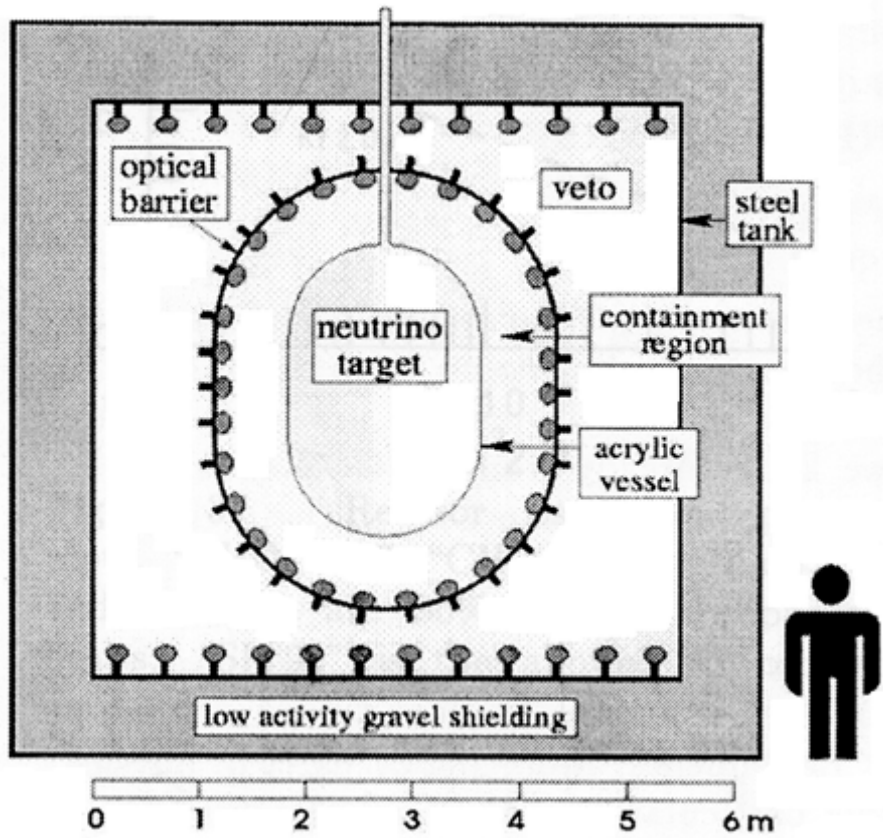
figure 1. The CHOOZ detector. \\

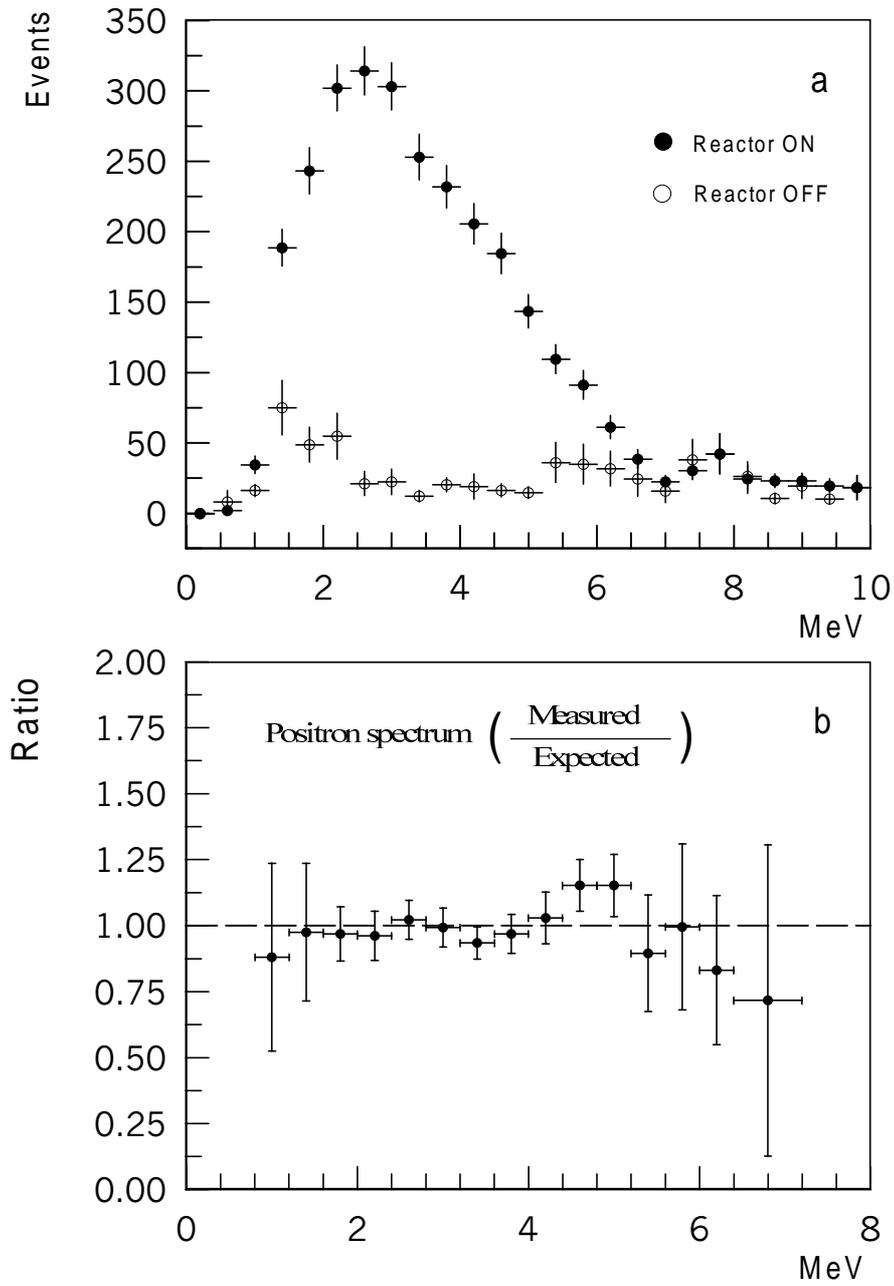
figure 2. CHOOZ positron energy spectrum 
a) spectra in reactor ON and OFF periods,
b) measured to expected no-oscillation ratio. \\

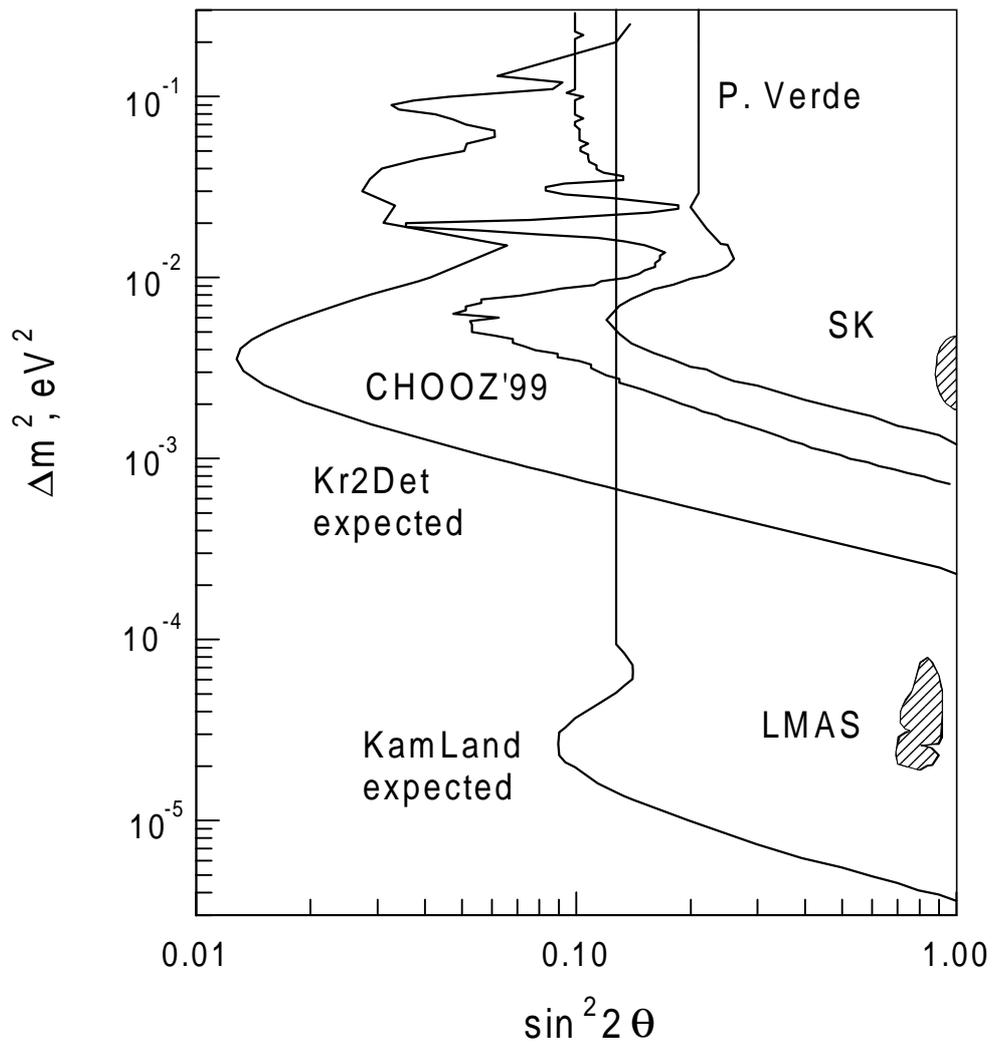
figure 3. Reactor neutrino (90\% CL) oscillation limits.
The shaded areas are the atmospheric and solar neutrino allowed oscillation regions.\\

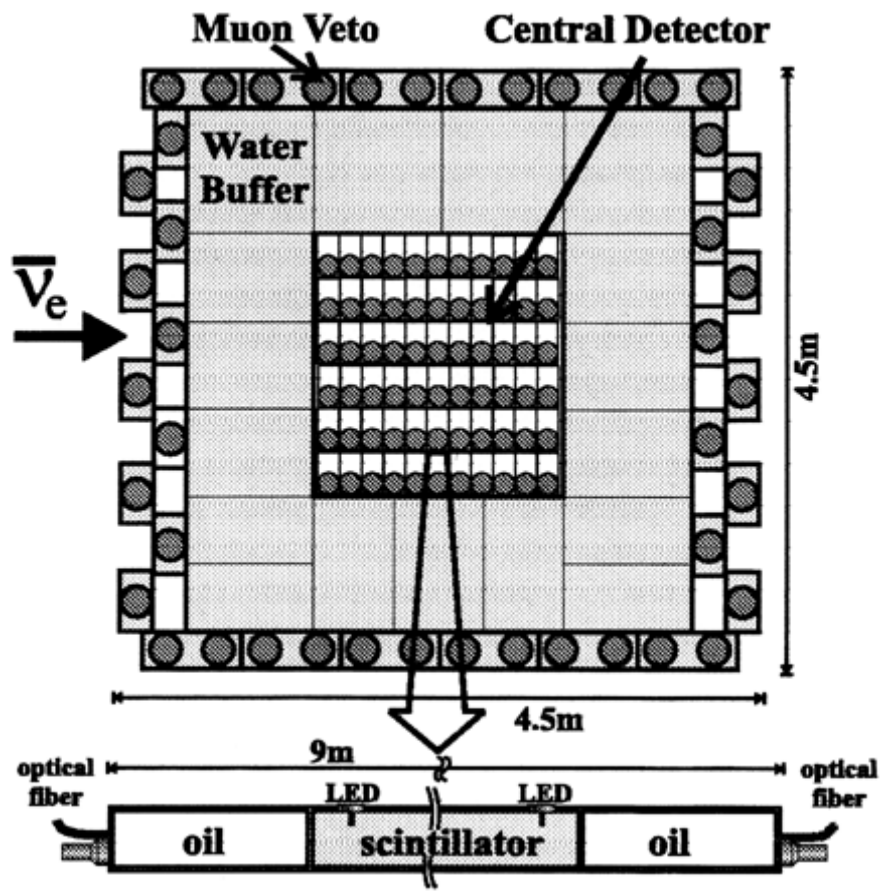
figure 4. The Palo-Verde detector.\\

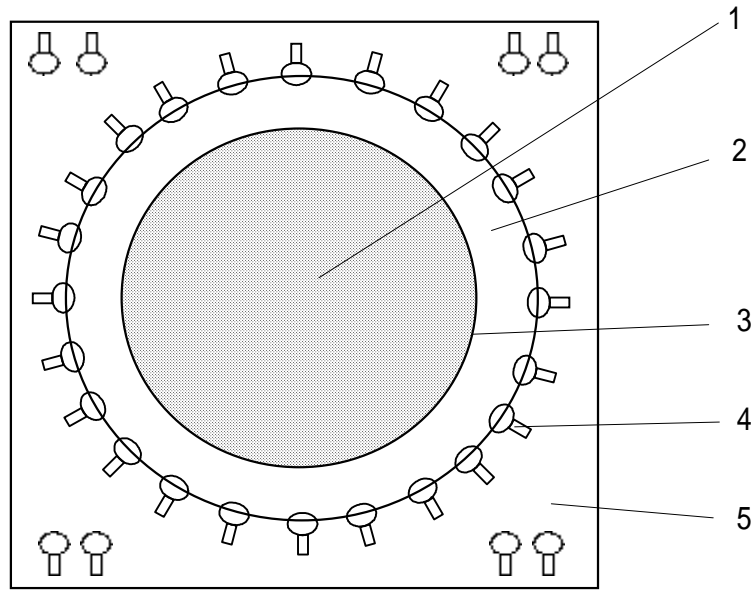
figure 5. The Kr2Det detector. \\
1 - The neutrino target (50 ton mineral oil + PPO),\\
2 - Mineral oil, 3 - The transparent film,\\
4 - The PMTs, 5 - Veto zone.\\

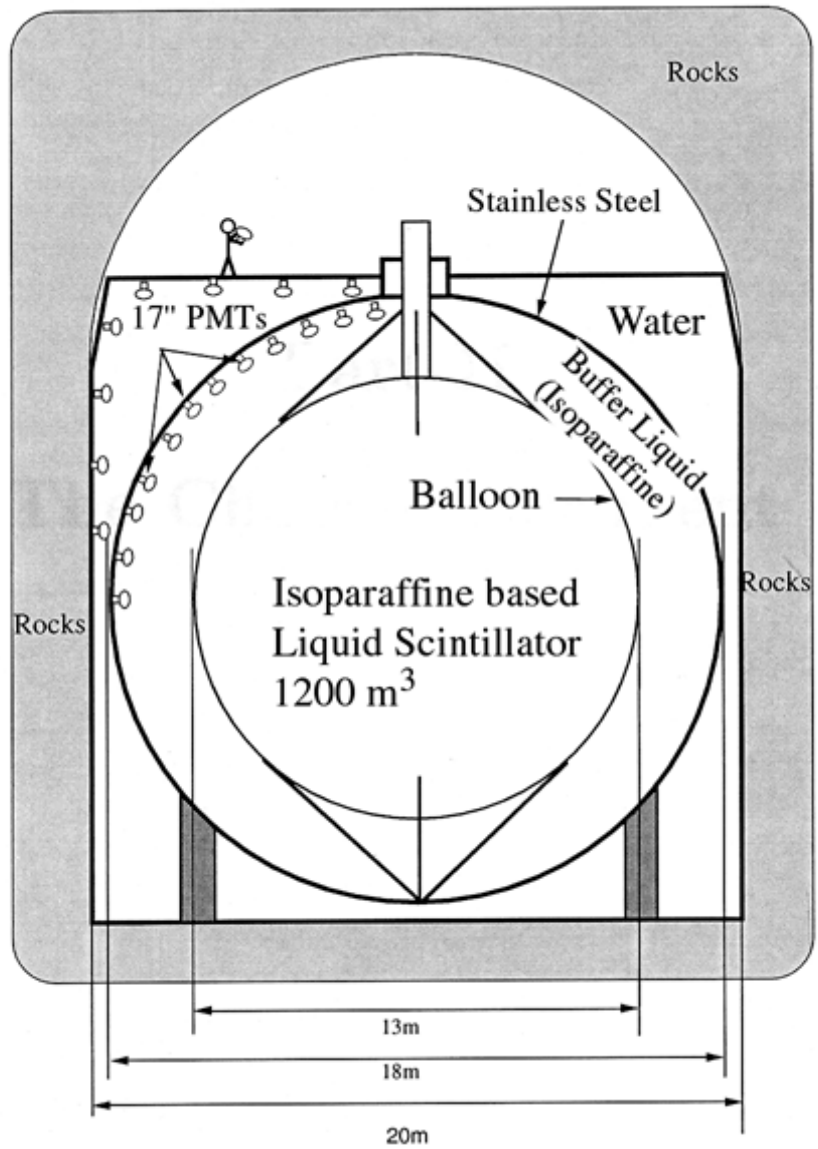
figure 6. The KamLAND detector.

\end{document}